\documentclass[12pt]{mynature}
\pdfoutput=1

\usepackage[english]{babel}
\usepackage{lineno, blindtext}
\usepackage{setspace}
\usepackage{amsmath}
\usepackage{amssymb}
\usepackage{caption}
\usepackage{graphicx}
\usepackage[utf8]{inputenc}
\usepackage[T1]{fontenc}
\usepackage[a4paper, margin=1in]{geometry}
\usepackage{float}
\typearea{12}

\makeatletter
\DeclareOldFontCommand{\rm}{\normalfont\rmfamily}{\mathrm}
\DeclareOldFontCommand{\sf}{\normalfont\sffamily}{\mathsf}
\DeclareOldFontCommand{\tt}{\normalfont\ttfamily}{\mathtt}
\DeclareOldFontCommand{\bf}{\normalfont\bfseries}{\mathbf}
\DeclareOldFontCommand{\it}{\normalfont\itshape}{\mathit}
\DeclareOldFontCommand{\sl}{\normalfont\slshape}{\@nomath\sl}
\DeclareOldFontCommand{\sc}{\normalfont\scshape}{\@nomath\sc}
\makeatother

\newcommand{\aap}{Astron. Astrophys.}
\newcommand{\aj}{Astron. J.}
\newcommand{\apj}{Astrophys. J.}

\newcommand{\apjs}{Astrophys. J. Suppl.}
\newcommand{\araa}{Annu. Rev. Astron. Astr.}

\newcommand{\jcap}{J. Cosmol. Astropart. P.}
\newcommand{\mnras}{Mon. Not. R. Astron. Soc.}
\newcommand{\nat}{Nature}
\newcommand{\pasp}{Publ. Astron. Soc. Pac.}
\newcommand{\physlettb}{Phys. Lett. B}
\newcommand{\physrep}{Phys. Rep.}
\newcommand{\physrev}{Phys. Rev.}

\newcommand{\rmxaa}{Rev. Mex. Astron. Astr.}
\newcommand{\sci}{Science}

\newcommand{\yp}{$Y_{\rm P}$}

\newcommand{\CIV}{C\,\textsc{iv}}
\newcommand{\DI}{\textrm{D\,{\sc i}}}
\newcommand{\HI}{\textrm{H\,{\sc i}}}
\newcommand{\HII}{\textrm{H\,{\sc ii}}}
\newcommand{\HeI}{\textrm{He\,{\sc i}}}

\newcommand{\Lya}{Ly$\alpha$}

\newcommand{\NHI}{$N(\textrm{H\,{\sc i}})$}
\newcommand{\NHeI}{$N(\textrm{He\,{\sc i}})$}

\newcommand{\NII}{N\,\textsc{ii}}
\newcommand{\SiII}{Si\,\textsc{ii}}

\newcommand{\qso}{HS\,1700$+$6416}

\newcommand{\zabss}{$z_{\rm abs}\simeq1.724$}

\DeclareCaptionFormat{figcapt}{{\bf Figure\!#1#2$\vert$} #3}
\DeclareCaptionFormat{tabcapt}{{\bf Table\!#1#2$\vert$} #3}
\DeclareCaptionFormat{suppfigcapt}{{\bf Supplementary Figure\!#1#2$\vert$} #3}
\DeclareCaptionFormat{supptabcapt}{{\bf Supplementary Table\!#1#2$\vert$} #3}
\usepackage{multibib}
\newcites{ap}{\LaTeX-Literature}
\begin{document}

\begin{LARGE}
\noindent

\textbf{Measurement of the primordial helium abundance from the intergalactic medium}
\end{LARGE}

~

\noindent
Ryan J. Cooke$^{1,2}$ \&\ Michele Fumagalli$^{1,3}$

\begin{footnotesize}
\noindent
$^{1}$Centre for Extragalactic Astronomy, Department of Physics, Durham University, South Road, Durham DH1 3LE, UK
\end{footnotesize}

\begin{footnotesize}
\noindent
$^{2}$Royal Society University Research Fellow
\end{footnotesize}

\begin{footnotesize}
\noindent
$^{3}$Institute for Computational Cosmology, Durham University, South Road, Durham DH1 3LE, UK
\end{footnotesize}

~

\textbf{Almost every helium atom in the Universe was created just a few minutes after the Big Bang through a process commonly referred to as Big Bang Nucleosynthesis\cite{AlpBetGam48,HoyTay64}. The amount of helium that was made during Big Bang Nucleosynthesis is determined by the combination of particle physics and cosmology\cite{SteSchGun77}. The current leading measures of the primordial helium abundance ($Y_{\rm P}$) are based on the relative strengths of \HI\ and \HeI\ emission lines emanating from star-forming regions in local metal-poor galaxies\cite{IzoThuGus14,AveOliSki15,PeiPeiLur16,FerTer18}. As the statistical errors on these measurements improve, it is essential to test for systematics by developing independent techniques. Here we report the first determination of the primordial helium abundance based on a near-pristine intergalactic gas cloud that is seen in absorption against the light of a background quasar. This gas cloud, observed when the Universe was just one-third of its present age ($z_{\rm abs} = 1.724$), has a metal content $\sim100$ times less than the Sun, and $\gtrsim30\%$ less metals than the most metal-poor \HII\ region currently known where a determination of the primordial helium abundance is afforded. We conclude that the helium abundance of this intergalactic gas cloud is $Y=0.250^{+0.033}_{-0.025}$, which agrees with the Standard Model primordial value\cite{Planck16,Pit18,Pat18}, $Y_{\rm P}=0.24672\pm0.00017$. Our determination of the primordial helium abundance is not yet as precise as that derived using metal-poor galaxies, but our method has the potential to offer a competitive test of physics beyond the Standard Model during Big Bang Nucleosynthesis.}

The near-pristine composition of some quasar absorption line systems\cite{CriOMeMur16,Leh16,CooPetSte17}, including systems with apparently pristine composition\cite{FumOMePro11}, makes intergalactic gas clouds well-suited for primordial abundance studies. Intergalactic gas clouds currently offer some of the most constraining observations of Big Bang Nucleosynthesis (BBN)\cite{CooPetSte18}, however, they have hitherto not been exploited to measure the primordial helium abundance. Here we present a novel measure of the helium abundance based on a gas cloud that is seen in absorption along the line-of-sight to the bright $z_{\rm em}=2.7348$ quasar \qso. This quasar intersects ten partial Lyman limit systems\cite{Fec06} (pLLSs; defined to have \HI\ column densities in the range $16.0\le\log N(\textrm{H\,\textsc{i}})/{\rm cm}^{-2}\le17.2$) and two of these systems are reported to have associated \HeI\ absorption based on low signal-to-noise and low resolution data acquired with the \emph{Hubble Space Telescope} (\emph{HST}) Faint Object Spectrograph (FOS)\cite{ReiVog93,SypShu13}.

We retrieved a High Resolution Echelle Spectrometer (HIRES) optical spectrum of \qso\ from the Keck Observatory Database of Ionized Absorption toward Quasars (KODIAQ)\cite{OMe15,OMe17}. These data cover the \HI\ \Lya\ absorption line and a selection of metal absorption lines. In order to obtain coverage across the \HI\ Lyman limit of the pLLS ($\sim912$\,\AA\ rest frame), we also retrieved the calibrated \emph{HST} FOS data of \qso\ from the \emph{HST} high level science products archive\cite{EvaKor04}. Finally, we retrieved the reduced data products of the \emph{HST} Cosmic Origins Spectrograph (COS) G130M observations (program 13491; PI, T. Tripp) of \qso\ from the Mikulski Archive for Space Telescopes, covering the \HeI\ resonant absorption lines of the pLLS at $z_{\rm abs}\simeq1.724$.

We have adopted an identical analysis procedure to that recently used to determine the primordial deuterium abundance with high precision\cite{Coo14, CooPetSte18}. We simultaneously fit the quasar emission and the absorption of the pLLS, while keeping blind the total column density of each ion (see Methods for further details). We uniquely determine the total \HI\ column density by fitting the flux decrement at the Lyman limit of the pLLS, modelling the quasar continuum as a power-law over the fitted range. The best-fitting model to the \HI\ Lyman limit absorption is shown in Figure~1a. We simultaneously fit several optically-thin high-order \HeI\ absorption lines, whose equivalent widths are directly proportional to the total \HeI\ opacity; we therefore uniquely determine the total \HeI\ column density, independent of the cloud model. The resulting profile fits are shown in Figure~1b. Finally, we detect a selection of weak metal lines associated with the pLLS, and we model these concurrently with the \HI\ and \HeI\ absorption lines (see Figure~2).

We stress that the details of the cloud model and component structure are not important since, for all of the observed ions, there is at least one weak absorption line available to uniquely determine the total column density of \HI, \HeI, and all of the observed metal species. Finally, we note that the entire fitting procedure is conducted in a blind fashion, such that the final parameter values are hidden from view until the analysis is complete. Once we had converged on the best-fitting profile model, the fitting procedure was not changed. The resulting values of the total column density are collected in Table~1. The final chi-squared of our best-fitting model is $\chi^{2}/{\rm dof}=2659.0/2354\simeq1.130$.

Based on this absorption line profile analysis, we measure \HI\ and \HeI\ column densities of
$\log_{10}\,(N({\rm H\,\textsc{i}})/{\rm cm}^{-2})=16.926\pm0.036$ and
$\log_{10}\,(N({\rm He\,\textsc{i}})/{\rm cm}^{-2}) = 15.851\pm0.055$, respectively,
corresponding to a neutral column density ratio \NHeI/\NHI~$= 0.085\pm0.013$. This measured ratio may be slightly different from the intrinsic He/H abundance ratio in a pLLS, depending on the incident radiation field. To convert the observed ratio into a determination of the helium number abundance, we construct a grid of \textit{Cloudy}\cite{Fer17} photoionisation simulations, where a gas slab is irradiated by the widely assumed ultraviolet background (UVB) of quasars and galaxies\cite{HarMad12}. This assumption is supported by the observation that there are no known bright galaxies in close proximity to the pLLS\cite{Sha05}. Our model grid covers a range of carbon metallicity ([C/H]; on a logarithmic scale relative to solar\cite{Asp09}), relative C/Si abundance ([C/Si]), total H volume density ($n_{\rm H}$), \NHI, and He/H abundance. Furthermore, we allow for some flexibility in the shape of the UVB by using a radiation field at an `effective redshift', $z_{\rm eff}$, in place of the observed redshift, \zabss.

We linearly interpolate our grid of simulations, and sample the parameter space using a Markov chain Monte Carlo procedure\cite{ForMac13} to identify the most likely set of parameters that reproduce the measured column densities. Probability distribution functions of the six model parameters are presented in Figure~3. The maximum likelihood parameter values are listed in the left column of Table 2. Based on these calculations, we find that the \zabss\ pLLS toward \qso\ has a volume density, [C/H] abundance, and [C/Si] relative abundance that are all typical of other pLLSs at this redshift\cite{Leh16}. We note that the metallicity of this absorption line system (${\rm [C/H]} = -1.688\pm0.059$ and ${\rm [Si/H]} = -1.969\pm0.088$, corresponding to $\sim1/50$ and $\sim1/100$ solar, respectively) is comparable to the most metal-poor \HII\ region\cite{IzoThu99} currently used to estimate \yp. The abundances we quote here represent the total gas phase abundance of these elements. Since pLLSs are believed to be dust-poor,\cite{Fum16} we do not need to apply a correction for dust.

The helium ionisation correction appears to be minimal in this system; the neutral column density ratio, \NHeI/\NHI~$= 0.085\pm0.013$, is consistent with the inferred abundance $y_{\rm P} \approx N$(He)/$N$(H)~$=0.085^{+0.015}_{-0.011}$. Furthermore, the quoted uncertainty is entirely dominated by the measurement uncertainty, and not by the errors associated with the ionisation modelling. We further explore the sensitivity of the ionisation correction to the assumed form of the incident radiation field in the Methods.

The helium number abundance, $y$, can also be expressed as a helium mass fraction, which is defined as
$Y \equiv 4\,y/(1+4\,y)$. Substitution of our MCMC samples of $y$ into this equation yields a mass fraction $Y = 0.250^{+0.033}_{-0.025}$ ($68\%~{\rm confidence}$),
constituting an $\sim11\%$ determination.
Our derived value of \yp\ is consistent with the Standard Model value\cite{Pit18} ($Y_{\rm P}=0.24672\pm0.00017$), assuming the \emph{Planck} baryon density\cite{Planck16} and the latest determination of the neutron lifetime\cite{Pat18}. Our measure is also in good agreement with the \emph{Planck} determination of the primordial helium abundance\cite{Planck16}, inferred from the damping tail of the CMB anisotropies (\yp\,$=0.253\pm0.021$; $68\%~{\rm confidence}$). However, at present, our measurement of $Y$ is not as precise as the value of $Y_{\rm P}$ derived from the CMB damping tail.

Most previous works have estimated \yp\ using observations of the most metal-poor \HII\ regions in the local universe\cite{IzoThuGus14,AveOliSki15,PeiPeiLur16,FerTer18}, and usually determine the primordial level by extrapolation to zero metallicity. Our novel measurement technique offers a new approach to pin down the primordial helium abundance in a near-pristine environment. Moreover, our determination is not impacted by the same systematics that currently affect the determination of \yp\ from metal-poor \HII\ regions\cite{IzoThuSta07}. The technique that we put forward in this paper offers an independent approach that will be critical to uncover potential systematics with both techniques.

Our analysis raises the exciting prospect of obtaining an independent high precision measure of the primordial helium abundance. Firstly, by collecting high-quality near-ultraviolet data around the \HI\ Lyman limit of the pLLS at \zabss\ toward \qso, it will be possible to pin down the total \HI\ column density to sub-percent level precision by fitting additional weak high-order \HI\ absorption lines jointly with the flux decrement at the Lyman limit; the current \NHI\ uncertainty is $\sim8\%$. Furthermore additional near ultraviolet data of this one sightline towards \qso\ would yield in excess of 10 independent He/H abundances.\cite{Fec06}

A more precise determination of \NHeI\ will require an absorption line system to be identified with more quiescent kinematics than the system analysed here. This will facilitate the detection of additional weak high order \HeI\ absorption lines. The identification of new \HeI\ absorbers should be considered a key goal for the future UV space-based observatories\cite{Mus18} that are on the horizon. \HeI\ absorbers will be easier to identify along sightlines\cite{OMe13} to quasars in the redshift interval $1.5\lesssim z_{\rm qso}\lesssim2.0$, where the contamination due to unrelated absorption near the high order \HI\ and \HeI\ absorption lines is considerably less compared with high redshift quasars.

The technique pioneered in this study could deliver a few per cent determination of \yp\ from near-pristine environments in the near-future. \yp\ offers a sensitive probe of the universal expansion rate at the time of BBN,\cite{SteSchGun77} and its measurement is therefore of acute interest in the quest to uncover deviations from the Standard Model of particle physics and cosmology.

\vspace{1.0cm}
\begingroup
\renewcommand{\section}[2]{}%

\endgroup

Correspondence and requests for materials should be addressed to R.J.C.\\
(ryan.j.cooke@durham.ac.uk).

\textbf{Acknowledgements}
We thank Gabor Worseck for a helpful discussion
about the \emph{HST}+COS data of this sightline.
During this work, R.~J.~C. was supported by a
Royal Society University Research Fellowship.
We acknowledge support from STFC (ST/L00075X/1, ST/P000541/1).
This project has received funding from the European Research Council 
(ERC) under the European Union's Horizon 2020 research and innovation 
programme (grant agreement No 757535).
This work used the DiRAC Data Centric system at Durham University,
operated by the Institute for Computational Cosmology on behalf of the
STFC DiRAC HPC Facility (www.dirac.ac.uk). This equipment was funded
by BIS National E-infrastructure capital grant ST/K00042X/1, STFC capital
grant ST/H008519/1, and STFC DiRAC Operations grant ST/K003267/1
and Durham University. DiRAC is part of the National E-Infrastructure.
This research has made use of NASA's Astrophysics Data System.

\textbf{Author Contributions}
Both authors participated in the interpretation and commented on the manuscript.
R.J.C. led the project and analysis, and was responsible for the text of the paper.
M.F. was responsible for the ionisation calculations.

\textbf{Author Information}
Reprints and permissions information is available at\\www.nature.com/reprints.
The authors declare no competing financial interests.
Readers are welcome to comment on the online version of the paper.

~

\begin{figure*}
  \centering
\vspace{-0.3cm}
 {\includegraphics[angle=0,height=190mm]{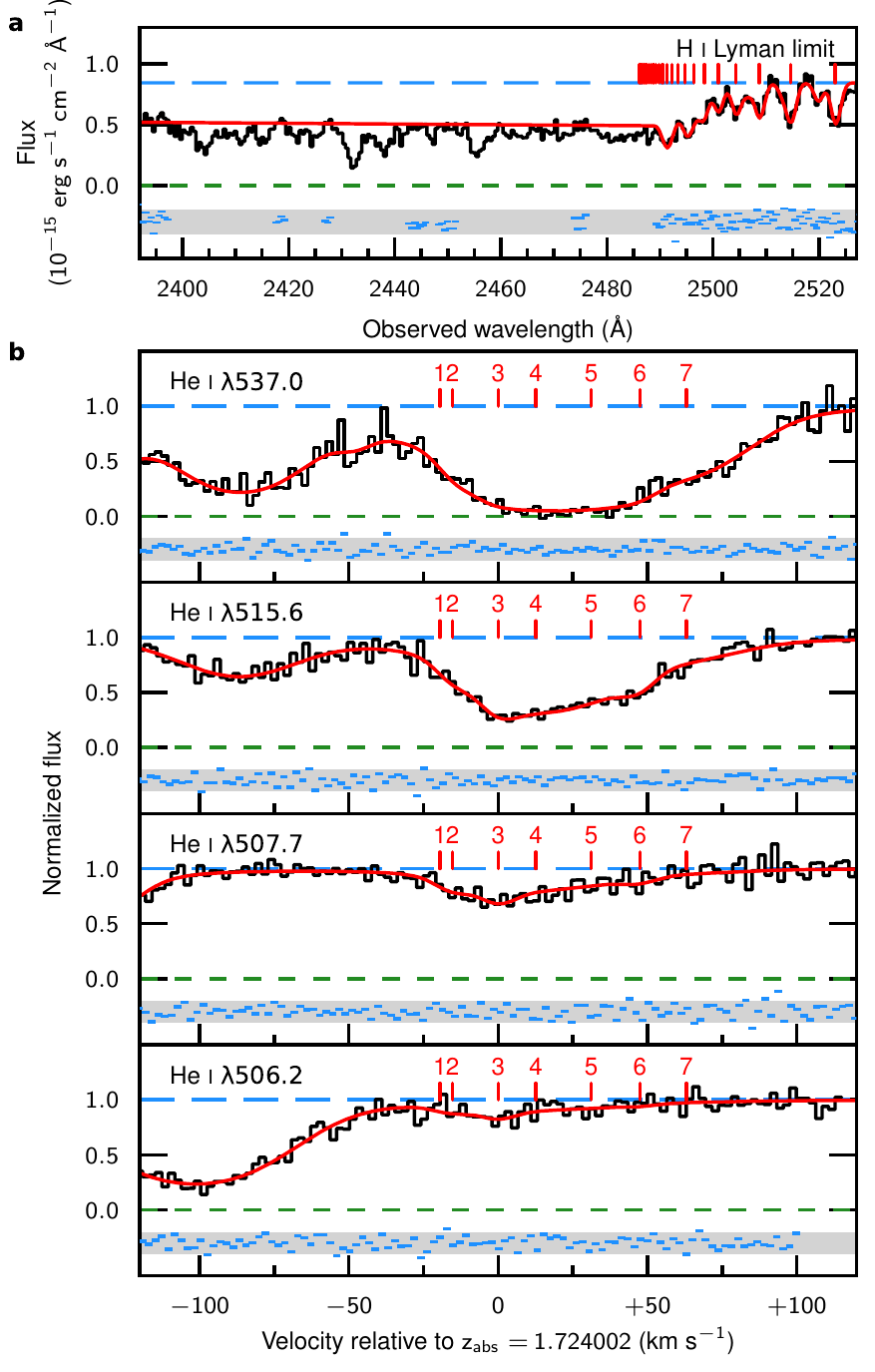}}\\
\vspace{-0.4cm}
  \caption{
\textbf{Hydrogen and helium absorption from an intergalactic gas cloud toward the quasar \qso.} \textbf{a,} High order \HI\ Lyman series absorption lines (indicated by the red tick marks above the spectrum) and \HI\ Lyman continuum absorption (blueward of an observed wavelength $\lambda\lesssim2483$\AA). The best-fit model is shown by the continuous red curve and the power-law fit to the quasar continuum is shown by the long dashed blue line. \textbf{b,} Four \HeI\ absorption lines associated with the pLLS at redshift \zabss. The red tick marks and associated numbers above the spectrum indicate the locations of the components comprising the absorption model. The blue points below all spectra are the normalised fit residuals, (data--model)/error, of all pixels that were used in the analysis, and the grey bands represent a confidence interval of $\pm2\sigma$. Labels at the top of all panels indicate the transition shown.
}
\end{figure*}

\begin{figure*}
  \centering
\vspace{-0.3cm}
 {\includegraphics[angle=0,height=190mm]{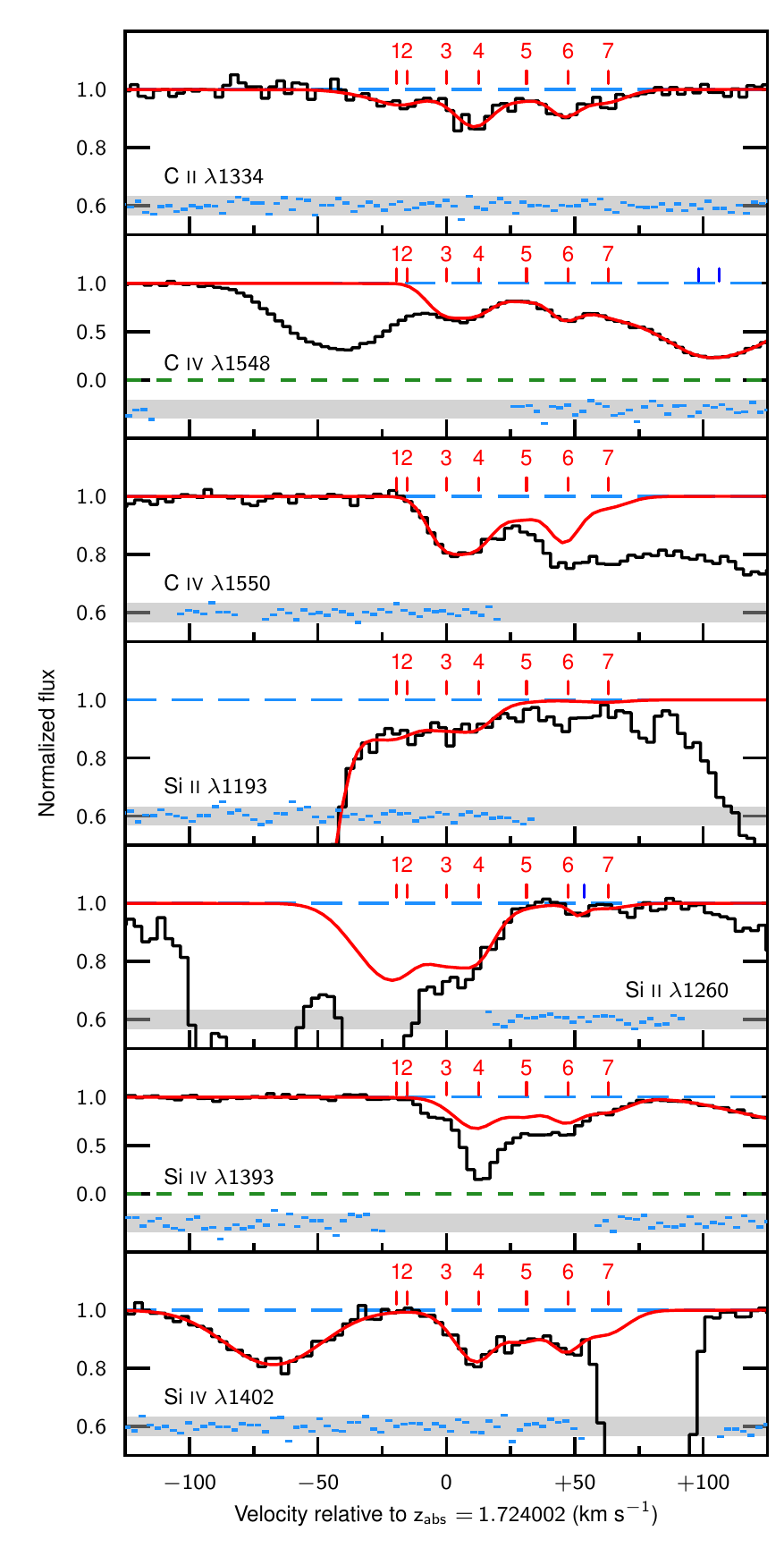}}\\
\vspace{-0.4cm}
  \caption{
\textbf{Heavy element absorption due to an intergalactic gas cloud toward the quasar \qso.} The red tick marks and associated numbers above each spectrum indicate the locations of the components comprising the absorption model. Fitted blends that are unrelated to the pLLS at \zabss\ are marked with blue tick marks above the spectrum. For example, the absorption feature at $v\simeq+100~{\rm km~s}^{-1}$ relative to \CIV\,$\lambda1548$ is an unrelated \Lya\ absorption line at $z_{\rm abs}=2.4703$, which is fully accounted for by simultaneously fitting the higher order \HI\ Lyman lines of this system. Similarly, the absorption feature at $v\simeq+50~{\rm km~s}^{-1}$ relative to \SiII\,$\lambda1260$ is an unrelated \NII\,$\lambda1083$ absorption line at $z_{\rm abs}=2.1679$. The blue points below all spectra are the normalised fit residuals, (data--model)/error, of all pixels that were used in our analysis, and the grey bands represent a confidence interval of $\pm2\sigma$. Note the different y-axis scale used on each panel.
}
\end{figure*}

\begin{figure*}
  \centering
\vspace{-0.3cm}
 {\includegraphics[angle=0,height=150mm]{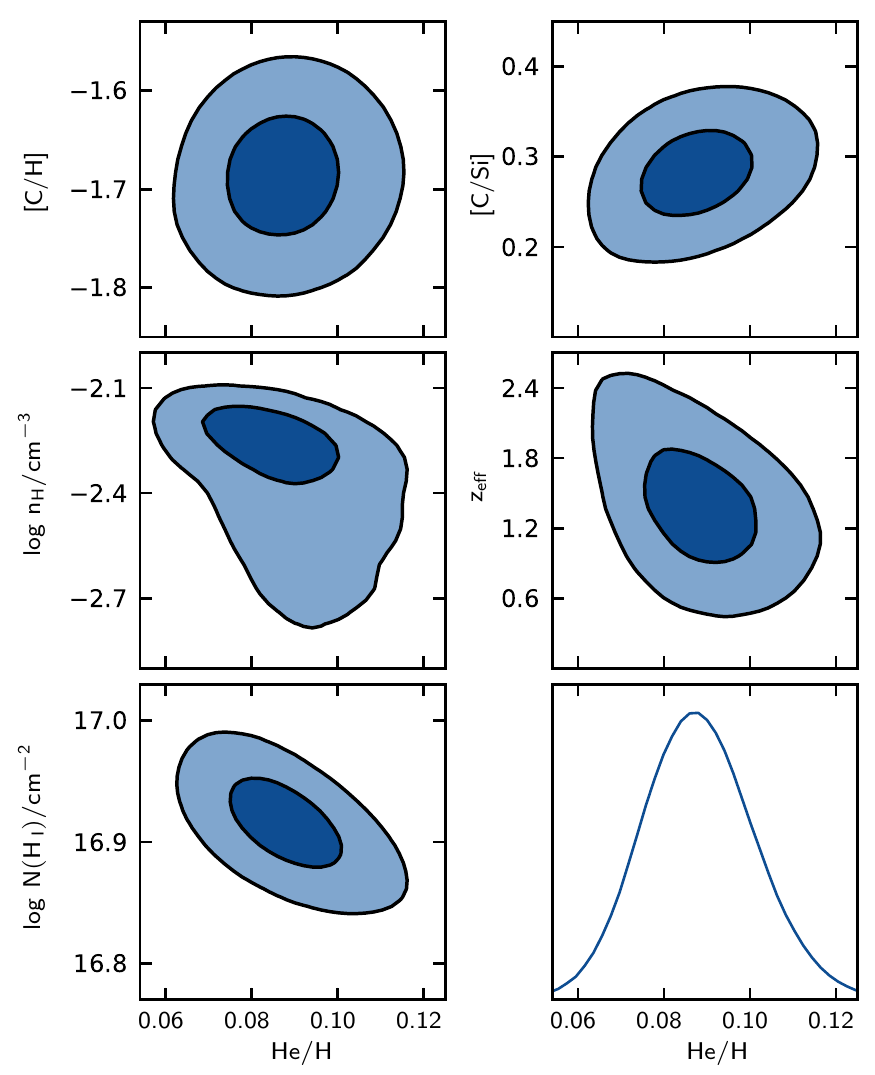}}\\
\vspace{-0.4cm}
  \caption{
\textbf{Confidence regions of the model parameters.} The 1D posterior distribution of He/H is shown in the bottom right panel. The remaining panels show the 68\%\ and 95\%\ confidence contours (dark and light shades, respectively) illustrating how each model parameter depends on the helium abundance.
}
\end{figure*}

\begin{table}[H]
    \centering
    \caption{\bf Ion column densities of the pLLS at \zabss\ toward \qso.}
    \label{tab:coldens}
    \begin{tabular}{llc}
        \hline
        Ion & Transitions used & log\,$N$(X)/cm$^{-2\,\,\dagger}$ \\
        \hline
        H\,\textsc{i}   & \Lya, Ly7--Lyman~Limit & $16.926\pm0.036$ \\
        He\,\textsc{i}  & $537.0, 515.6, 507.7, 506.2$ & $15.851\pm0.055$ \\
        C\,\textsc{ii}  & $1334$ & $13.189\pm0.043$ \\
        C\,\textsc{iv}  & $1548, 1550$ & $13.446\pm0.047$ \\
        Si\,\textsc{ii} & $1193, 1260$ & $12.586\pm0.059$ \\
        Si\,\textsc{iv} & $1393, 1402$ & $13.002\pm0.028$ \\
        \hline
    \end{tabular}

    $^{\dagger}$\,Quoted uncertainties represent the 68 per cent confidence interval.
\end{table}

\begin{table}[H]
    \centering
    \caption{\bf Best-fitting parameter values of our ionisation modelling.}
    \label{tab:params}
    \begin{tabular}{lcc}
        \hline
        Parameter & UVB only & UVB+galaxy \\
        \hline
        $\log\,N({\rm H\,\textsc{i}})$   & $16.908^{+0.039}_{-0.034}$ & $16.911^{+0.038}_{-0.035}$ \\
        $\log\,n_{\rm H}$  & $-2.28^{+0.08}_{-0.24}$ & $-2.23^{+0.10}_{-0.03}$ \\
        $z_{\rm eff}$  & $1.21^{+0.60}_{-0.38}$ & $\ldots$ \\
        $\log\,j_{\rm gal}$  & $\ldots$ & $\le-0.45^{\dagger}$ \\
        ${\rm [C/H]}$  & $-1.696^{+0.063}_{-0.055}$ & $-1.732^{+0.075}_{-0.229}$ \\
        ${\rm [C/Si]}$ & $+0.255^{+0.060}_{-0.037}$ & $+0.255^{+0.052}_{-0.055}$ \\
        ${\rm He/H}$ & $0.085^{+0.015}_{-0.011}$ & $0.077^{+0.015}_{-0.037}$ \\
        \hline
    \end{tabular}

    $^{\dagger}$\,95\%~{\rm Confidence~Interval}
\end{table}

\clearpage
\newpage

\textbf{Methods}

\textbf{Observational Data.} The $z_{\rm em}=2.7348$ quasar \qso\ was first identified as part of the Hamburg quasar survey\cite{Rei89}. Since its discovery, \qso\ has been observed multiple times, and across many wavebands. We utilise archival data taken with the W.~M.~Keck Observatory High Resolution Echelle Spectrometer (HIRES) in addition to archival data acquired with \emph{HST} FOS and \emph{HST} COS.

\textbf{HIRES data.} Using custom built routines, we combined all HIRES data from the KODIAQ\cite{OMe15,OMe17} database that were acquired through decker C1 ($0.861''\times7.0''$) over the lifetime of HIRES, amounting to a total exposure time of 30,417\,s. The C1 decker has a nominal spectral resolution of $R~=48,000$, corresponding to a velocity full-width at half-maximum (FWHM) resolution of $v_{\rm FWHM}=6.25~{\rm km~s}^{-1}$. The combined data cover a wavelength range 3109--6076\,\AA, which includes \HI\ \Lya\ and a selection of metal absorption lines. The final combined spectrum has a signal-to-noise ratio per pixel of S/N~$\sim30$ near \Lya\ and S/N~$\sim50$ near the metal absorption lines.

\textbf{HST FOS data.} We retrieved the fully reduced, coadded and calibrated \emph{HST} FOS data of \qso\ from the \emph{HST} high level science products archive\cite{EvaKor04}. The full data cover the wavelength range 1140--3277\AA, with a spectral resolution $R\simeq1,300$. Near the \HI\ Lyman limit, the data exhibit a signal-to-noise ratio per pixel of S/N~$\simeq25$.

\textbf{HST COS data.} We retrieved the final calibrated data products of the \emph{HST} COS G130M observations of \qso\ from the Mikulski Archive for Space Telescopes. The processed \textsc{x1dsum} files were combined using version 2.1 of the \textsc{coadd\_x1d} software package. These high quality COS data cover the \HeI\ resonant absorption lines of the pLLS at \zabss. The final combined signal-to-noise ratio per $2.2~{\rm km~s}^{-1}$ pixel of the data near \HeI\,$\lambda507.7$ is S/N~$\simeq20$.

\textbf{Atomic Data.} The atomic data used in our analysis were retrieved from several sources. For the \HI\ and metal absorption lines, we use the atomic data compiled by Morton\cite{Mor03}. In order to fit the \HI\ Lyman limit absorption, we take the photoionisation cross-section of \HI\ from the Verner et al. database\cite{Ver96}. Finally, we retrieved all of the \HeI\ vacuum wavelengths ($\lambda_{0}$) and oscillator strengths ($f$) from the National Institutes of Standards and Technology (NIST) Atomic Spectral Database (ASD)\cite{NISTASD}. We computed the associated natural damping constant ($\gamma_{\rm ul}$) of all transitions by summing the spontaneous transition probabilities to all lower levels
\begin{equation}
\gamma_{\rm ul}=\sum\limits_{l} A_{\rm ul}
\end{equation}
All of the relevant values are listed in Supplementary Table 1.

The pLLS at \zabss\ toward the quasar \qso\ is detected in \HeI\ absorption from $\lambda537.0$ to $\lambda505.5$. However, the $f$-values and damping constants of these \HeI\ absorption lines are only available for wavelengths longward of \HeI\,$\lambda507.0$; the atomic data (both $f$ and $\gamma_{\rm ul}$) of the higher order \HeI\ lines are not available in the NIST ASD. To overcome this limitation, we fit a low-order polynomial to the available atomic data and extrapolate the polynomial fit to estimate the unknown values. We use a third-order polynomial for the $f$-values and a second-order polynomial for the damping constants. To test the accuracy of this extrapolation, we applied an identical procedure to the atomic data of neutral hydrogen, by fitting the atomic data of \Lya-Ly9 and compare the extrapolated values of Ly10-Ly14 to the true values reported by Morton\cite{Mor03}. We found that this simple extrapolation allows us to predict the \HI\ Ly10-Ly14 $f$-values to better than $2\%$ accuracy, and the damping constants to within $25\%$ accuracy.

The only extrapolated value that is used in our analysis is that of \HeI\,$\lambda506.2$. Since this line is weak, the accuracy of the damping constant is not important. However, the total \HeI\ column density is degenerate with the $f$-value. We estimate that the systematic uncertainty of the \HeI\,$\lambda506.2$~$f$-value to be $\lesssim0.5\%$, which translates to a systematic limit of $0.5\%$ on the total \HeI\ column density. As shown below, this level of precision is well below the measurement precision of the \HeI\ column density.

\textbf{Fitting Procedure.} We use the Absorption LIne Software (\textsc{alis}) to derive the column densities of the absorption lines associated with the pLLS at \zabss\ toward \qso. \textsc{alis} uses a chi-squared minimisation algorithm to determine the model parameters that best fit the data, weighted by the inverse variance of each pixel that is used in the analysis.

We perform a simultaneous fit to the quasar emission and the absorption of the pLLS. This ensures that any uncertainty associated with the emission profile is folded into the ion column density uncertainties. We model the quasar emission by a low-order Legendre polynomial locally around each absorption line. Typically, we use a polynomial of degree $\lesssim4$ to describe the quasar continuum. All absorption lines are modelled as a Voigt profile, which is characterised by a column density, a redshift, and a Doppler parameter. Since the pLLS studied here contains several absorption components, we fit directly for the total column density of each ion. For all fitted ions, we only include pixels in our analysis that we deem are free from unrelated absorption. In some cases, we identify blends and fully account for these by simultaneously fitting a Voigt profile. Finally, we include two global model parameters to define the zero level of the HIRES and COS data, to account for any small residual to the background subtraction.

For the \HI\ \Lya\ profile, we include \DI\ absorption at the primordial ratio\cite{CooPetSte18} (i.e. $N$(\DI)\,= \NHI$\,-\,4.5974$ for each component of \HI), and several unrelated blends. Since the equivalent width of the \HI\ \Lya\ absorption line lies on the flat part of the curve of growth, the column density is degenerate with the cloud model. However, the total \HI\ column density is uniquely determined by fitting the flux decrement at the Lyman limit of the pLLS. The optical depth beyond the \HI\ Lyman limit is modelled as $\tau_{\nu}=\,$\NHI\,$\sigma_{\nu}$, where $\sigma_{\nu}$ is the photoionisation cross-section of \HI.

In order to fit the observed component structure of the metal absorption lines, we require seven absorption components. The redshift and Doppler parameter of each component are allowed to vary, but are forced to be the same across all species. We only include pixels in the metal absorption line fitting that we deem to be free of unrelated absorption. Where possible, we account for line blending due to unrelated absorption features provided that the blend can be fully accounted for using another transition. For example, the absorption feature at $v\simeq+100~{\rm km~s}^{-1}$ relative to \CIV\,$\lambda1548$ corresponds to \Lya\ absorption at $z_{\rm abs}=2.4703$, which we account for by simultaneously fitting the corresponding higher order Lyman series lines.

We use the same component redshifts to fit the \HeI\ absorption lines, but allow the Doppler parameter of each line to vary independently of the metal absorption lines. In our analysis, we only include the \HeI\ absorption lines that are entirely free of line blending, to reduce the risk of contamination from the rich forest of metal absorption lines\cite{Fec06} associated with absorption systems in the range $0.2<z<2.6$. The spectra near all of the available \HeI\ resonant absorption lines are overlaid with the best-fitting model in Supplementary Figure 1.

\textbf{Ionisation Modelling.} Since the gas associated with pLLSs is mostly ionised, both \HeI\ and \HI\ are minor ions; thus, \HeI\ may not precisely trace the \HI\ gas, and a small ionisation correction may be required to convert the measured $N$(\HeI)/$N$(\HI) ratio into a He/H abundance.

To estimate the magnitude of this correction, we use version 17.0 of the \textit{Cloudy} photoionisation simulation software\cite{Fer17} to model the pLLS as a slab of constant density gas in ionisation and thermal equilibrium, irradiated by the CMB and ultraviolet background (UVB) of galaxies and quasars. The UVB radiation field at a given redshift is defined by a shape and intensity (see Supplementary Figure~2 for five example spectra). We allow for some flexibility in the shape of the UVB by irradiating the pLLS by the widely-assumed Haardt \&\ Madau\cite{HarMad12} UVB at an `effective redshift' $z_{\rm eff}$; we do not enforce a prior of the UVB based on the measured redshift, $z_{\rm abs}$. The incident radiation field of our fiducial model is therefore $J(\nu) = J_{\rm CMB}(\nu, z_{\rm abs}) + J_{\rm HM12}(\nu, z_{\rm eff})$.

We consider a grid of UVB radiation fields between $0 \le z_{\rm eff} \le 6$, sampled in steps of $\Delta z_{\rm eff} = 0.25$ when $z_{\rm eff} < 2$ and in steps of $\Delta z_{\rm eff} = 0.5$ when $z_{\rm eff} > 2$. These UVB models set both the intensity and shape of the radiation field; to allow for some flexibility in the ionisation parameter at a given redshift, we construct a grid of models to cover a range of volume density ($-4.6<\log_{10}\,(n_{\rm H}/{\rm cm}^{-3})<+0.2$, in steps of 0.2~dex).
Note that the ionisation parameter, which is defined as the ratio of gas number density to photon number density, is what determines the ionisation state of the gas.

Our grid of \textit{Cloudy} models also includes a range of metallicity ($-4.0<{\rm [X/H]}<-0.8$, in steps of 0.2~dex),
\HI\ column density ($16.6 < \log_{10}\,(N(\textsc{H\,\textsc{i}})/{\rm cm}^{-2}) < 17.2$, in steps of 0.15~dex), and
helium abundance ($0.05 < ({\rm He/H})/0.084249 < 1.5$, in linear steps of 0.05). Grains are not included in our calculations.

Our \textit{Cloudy} models are stopped once the input neutral hydrogen column density is reached, at which point we store the column densities of each ion stage of H, He, C, and Si. Our grid of models is linearly interpolated and sampled using a Markov chain Monte Carlo procedure\cite{ForMac13} to identify the parameter values that reproduce the measured column densities listed in Table~1 (see Supplementary Figure~3). Our model contains six parameters (He/H, $z_{\rm eff}$, $n_{\rm H}$, [C/H], [C/Si], and \NHI), which are exactly determined by the six reported column density measurements. The resulting parameter values, together with their $68\%$ confidence intervals, are listed in the penultimate column of Table~2.

One assumption of our fiducial model is that the incident radiation field is that of the metagalactic UVB calculated by Haardt \&\ Madau\cite{HarMad12}. However, due to the uncertainty of the UVB\cite{FacGig09} and the fact that some pLLSs arise in the outskirts of a nearby galaxy\cite{Fum11}, we have tested the sensitivity of the helium ionisation correction to the assumed radiation field. In what follows, we assume that the incident radiation field comprises the CMB, the Haardt \&\ Madau\cite{HarMad12} UVB at a redshift \zabss, and a galaxy spectrum, $J_{\rm G}(\nu)$. This is similar to an approach recently used to model Lyman Limit Systems\cite{Fum16}. The incident radiation field in this case is of the form $J(\nu) = J_{\rm CMB}(\nu, z_{\rm abs}) + J_{\rm HM12}(\nu, z_{\rm abs}) + j_{\rm gal}\,J_{\rm G}(\nu)$, where $j_{\rm gal}$ sets the intensity of the galactic spectrum at the surface of the pLLS, and $J_{\rm G}(\nu)$ is defined such that when $j_{\rm gal}=1$, it has the same intensity of H Lyman continuum photons as $J_{\rm HM12}(\nu, z_{\rm abs})$. We define the spectral shape of $J_{\rm G}(\nu)$ by synthesising a \textsc{starburst99} model\cite{Lei99} (v7.0.1) using the default input parameters and assuming a continuous star formation rate of $\dot{\psi}=1\,M_{\odot}~{\rm yr}^{-1}$. Our galaxy model employs the Geneva (2012)\cite{Eks12} solar metallicity stellar models, without rotation. The galactic radiation field, when added to the UVB, results in a softer spectrum relative to the UVB alone; we illustrate five example spectra of $J(\nu)$ in Supplementary Figure~4.

With this alternative definition of $J(\nu)$, we generated a grid of \textit{Cloudy} models that are otherwise identical to that described earlier, and sampled the grid with a Markov chain Monte Carlo procedure (see Supplementary Figure~5). The maximum likelihood parameter values, along with their corresponding $68\%$ confidence intervals, are listed in the final column of Table~2. These parameter values are all within $1\sigma$ of our fiducial model. The known galaxy population\cite{Sha05} along this sightline consists of four galaxies with redshifts that are consistent with \zabss. The closest galaxy (BX\,756) is $\sim500~{\rm kpc}$ in projection from the pLLS. The remaining three galaxies are $\gtrsim1~{\rm Mpc}$ in projection from the pLLS. Thus there does not appear to be a significant bright galaxy population in the immediate vicinity of the pLLS. This is consistent with the maximum likelihood value $j_{\rm gal}=0$, implying that the radiation field incident on the pLLS at \zabss\ is likely dominated by the metagalactic UVB. We therefore conclude that our assumed shape for $J(\nu)$ in the fiducial model is supported by the data in hand.

By changing the form of the incident radiation field, the central He/H values of our two models differ by $\sim9\%$ (compare the values listed in the final row of Table~2). Thus, the shape of the incident radiation field must be known before a robust ionisation correction can be applied. In principle, the shape of the radiation field can be inferred by measuring the column density ratio of successive ions of the same element, particularly the ions that bracket the ionisation potentials of H and He. Thus, in order to pin down the radiation field with the accuracy demanded by precision cosmology, it will be important to reliably determine the column densities of metal absorption lines from a range of ionisation stages.

\textbf{Data availability.}
The data that support the plots within this paper and other findings of this study are available from the corresponding author upon reasonable request.
The Keck HIRES data are available as a high level science product, housed at the Keck Observatory Data Archive, which is available from\\
https://koa.ipac.caltech.edu/cgi-bin/KOA/nph-KOAlogin.\\
The fully reduced \emph{HST} FOS data are available from\\
https://archive.stsci.edu/prepds/fos\_agn/.\\
The reduced \emph{HST} COS data are available at\\
http://archive.stsci.edu/proposal\_search.php?mission=hst\&id=13491.


\vspace{1.0cm}
\begingroup
\renewcommand{\section}[2]{}%

\endgroup


\newpage

\captionsetup[figure]{format=suppfigcapt,labelfont={bf},name={},labelsep=space}
\captionsetup[table]{format=supptabcapt,labelfont={},name={},labelsep=space}
\setcounter{figure}{0}
\setcounter{table}{0} 

\begin{table}
    \centering
    \caption{\bf Atomic data of the \HeI\ resonant absorption lines.}
    \label{tab:atomic}
    \begin{tabular}{lccc}
        \hline
        Ion & $\lambda_{0}$ & $f$ & $\gamma_{\rm ul}$\\
            & (\AA)     &     & $({\rm s}^{-1})$\\
        \hline
        He\,\textsc{i} $\lambda584.3$ & $584.3344$ & $0.2763$ & $1.799{\rm E}9$\\
        He\,\textsc{i} $\lambda537.0$ & $537.0299$ & $0.07346$ & $5.799{\rm E}8$\\
        He\,\textsc{i} $\lambda522.2$ & $522.2131$ & $0.02988$ & $2.520{\rm E}8$\\
        He\,\textsc{i} $\lambda515.6$ & $515.6168$ & $0.01504$ & $1.308{\rm E}8$\\
        He\,\textsc{i} $\lambda512.1$ & $512.0986$ & $0.008630$ & $7.634{\rm E}7$\\
        He\,\textsc{i} $\lambda510.0$ & $509.9983$ & $0.005407$ & $4.939{\rm E}7$\\
        He\,\textsc{i} $\lambda508.6$ & $508.6434$ & $0.003611$ & $3.248{\rm E}7$\\
        He\,\textsc{i} $\lambda507.7$ & $507.7181$ & $0.002531$ & $2.287{\rm E}7$\\
        He\,\textsc{i} $\lambda507.1$ & $507.0580$ & $0.001842$ & $1.670{\rm E}7$\\
        He\,\textsc{i} $\lambda506.6$ & $506.5705$ & $0.001382^{\dagger}$ & $1.221{\rm E}7^{\dagger}$\\
        He\,\textsc{i} $\lambda506.2$ & $506.2003$ & $0.001064^{\dagger}$ & $9.247{\rm E}6^{\dagger}$\\
        He\,\textsc{i} $\lambda505.9$ & $505.9125$ & $0.0008363^{\dagger}$ & $7.143{\rm E}6^{\dagger}$\\
        He\,\textsc{i} $\lambda505.7$ & $505.6843$ & $0.0006693^{\dagger}$ & $5.609{\rm E}6^{\dagger}$\\
        He\,\textsc{i} $\lambda505.5$ & $505.5004$ & $0.0005440^{\dagger}$ & $4.469{\rm E}6^{\dagger}$\\
        \hline
    \end{tabular}

$^{\dagger}${No atomic data exist for these transitions. Instead, these values were extrapolated based on a polynomial fit to the known measures from lower order lines (see Methods).}\\
\end{table}

\begin{figure*}
  \centering
\vspace{-0.3cm}
 {\includegraphics[angle=0,height=200mm]{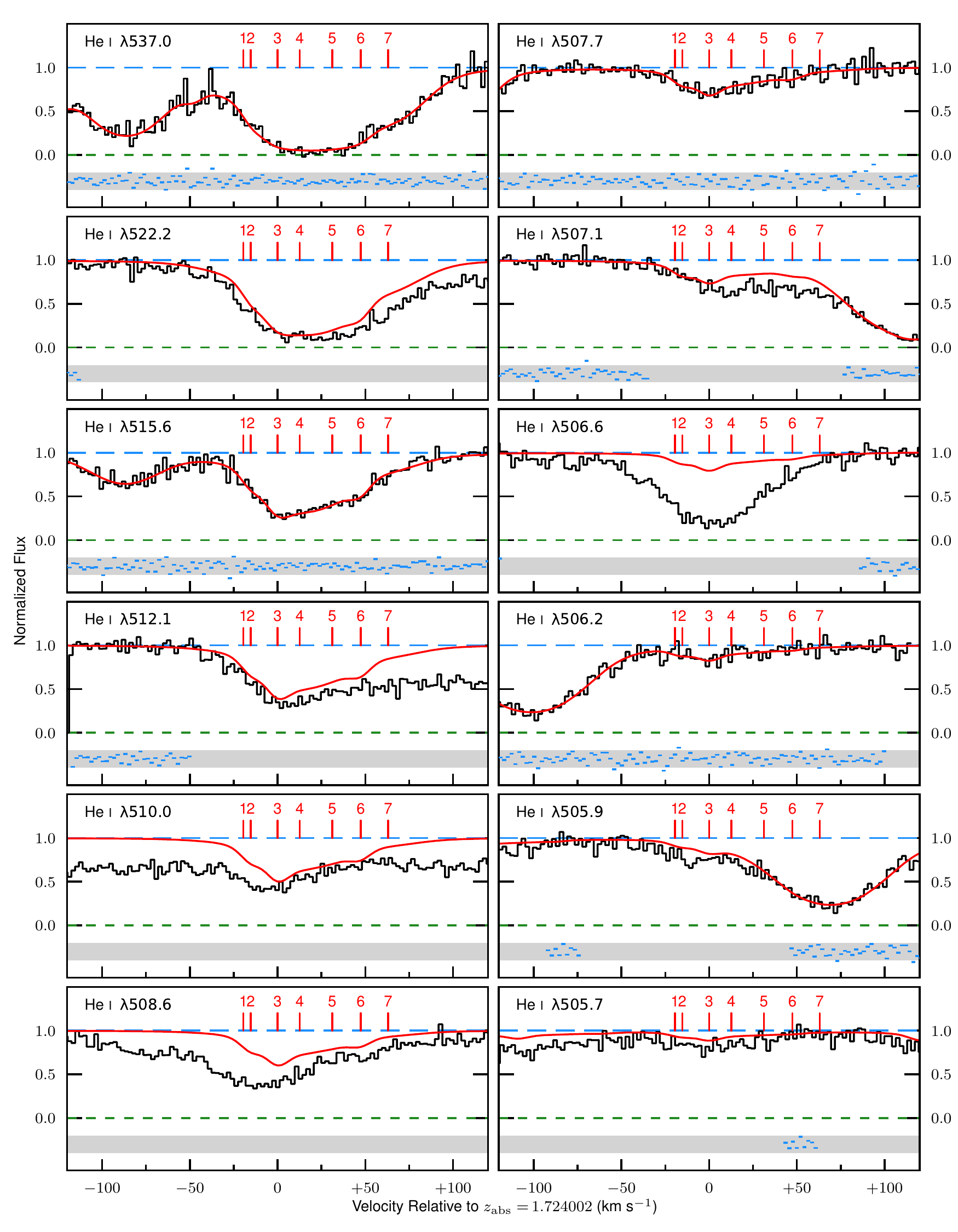}}\\
\vspace{-0.4cm}
  \caption{
\textbf{Resonant absorption lines of neutral helium from an intergalactic gas cloud at \zabss\ toward the quasar \qso.} The red curve shows the best-fitting model profile. The data and model are normalised to the best-fitting continuum model (long dashed blue line) and corrected for the fitted zero-level (short dashed green line). The red tick marks and associated numbers above the spectrum indicate the locations of the components comprising the absorption model.  The \HeI\ absorption lines that are used to pin down the total \HeI\ column density include: \HeI\,$\lambda537.0, \lambda515.6, \lambda507.7,$ and $\lambda506.2$. The blue points below all spectra are the normalised fit residuals, (data--model)/error, of all pixels that were used in the analysis, and the grey bands represent a confidence interval of $\pm2\sigma$.
A label in the top left corner of every panel indicates the \HeI\ absorption line shown.
}
  \label{fig:HeI}
\end{figure*}

\begin{figure}
  \centering
 {\includegraphics[angle=0,width=150mm]{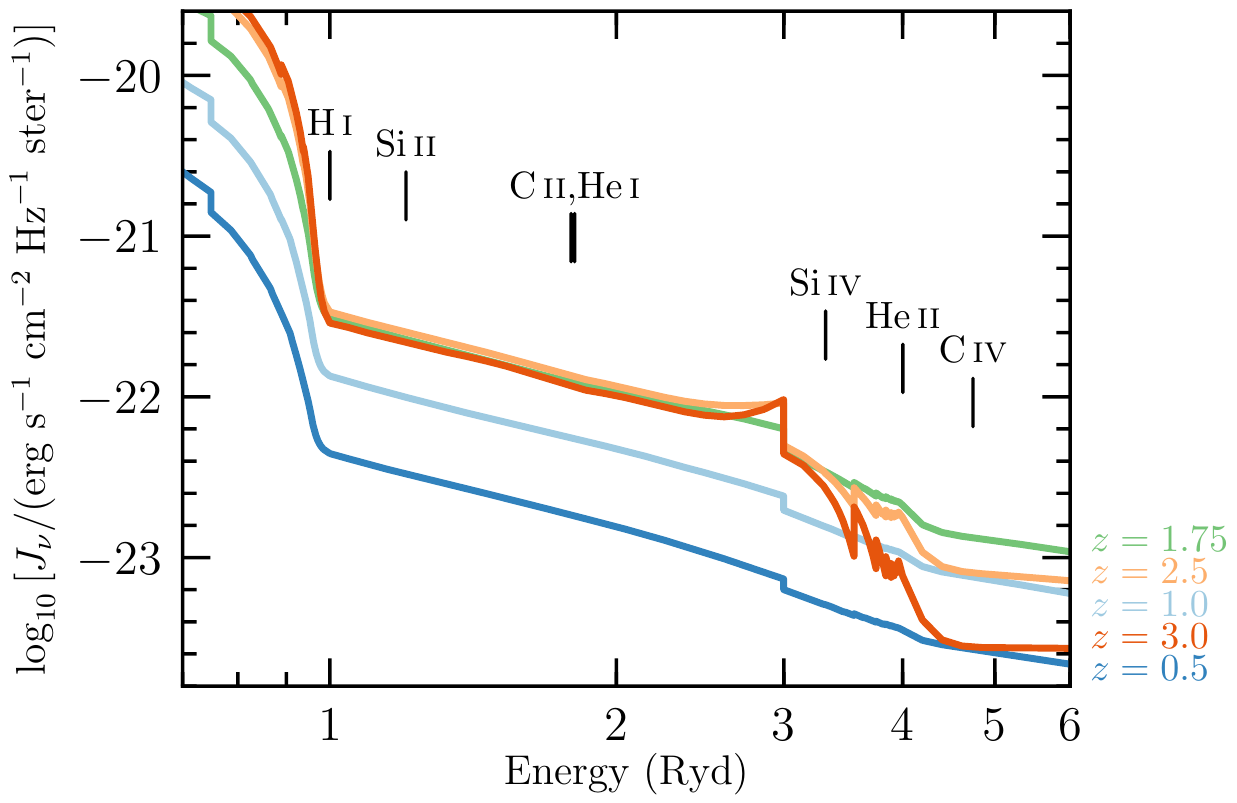}}\\
  \caption{
\textbf{Five sample spectra illustrate how the Haardt \&\ Madau UVB radiation field varies as a function of redshift.} The redshift corresponding to each spectrum is indicated at the bottom right of the figure. Labelled tick marks above the spectra indicate the ionisation potentials of the relevant ions that are used in this letter.
}
  \label{fig:uvb}
\end{figure}

\begin{figure}
\noindent\makebox[\textwidth][c]{\begin{minipage}{183mm}
\includegraphics{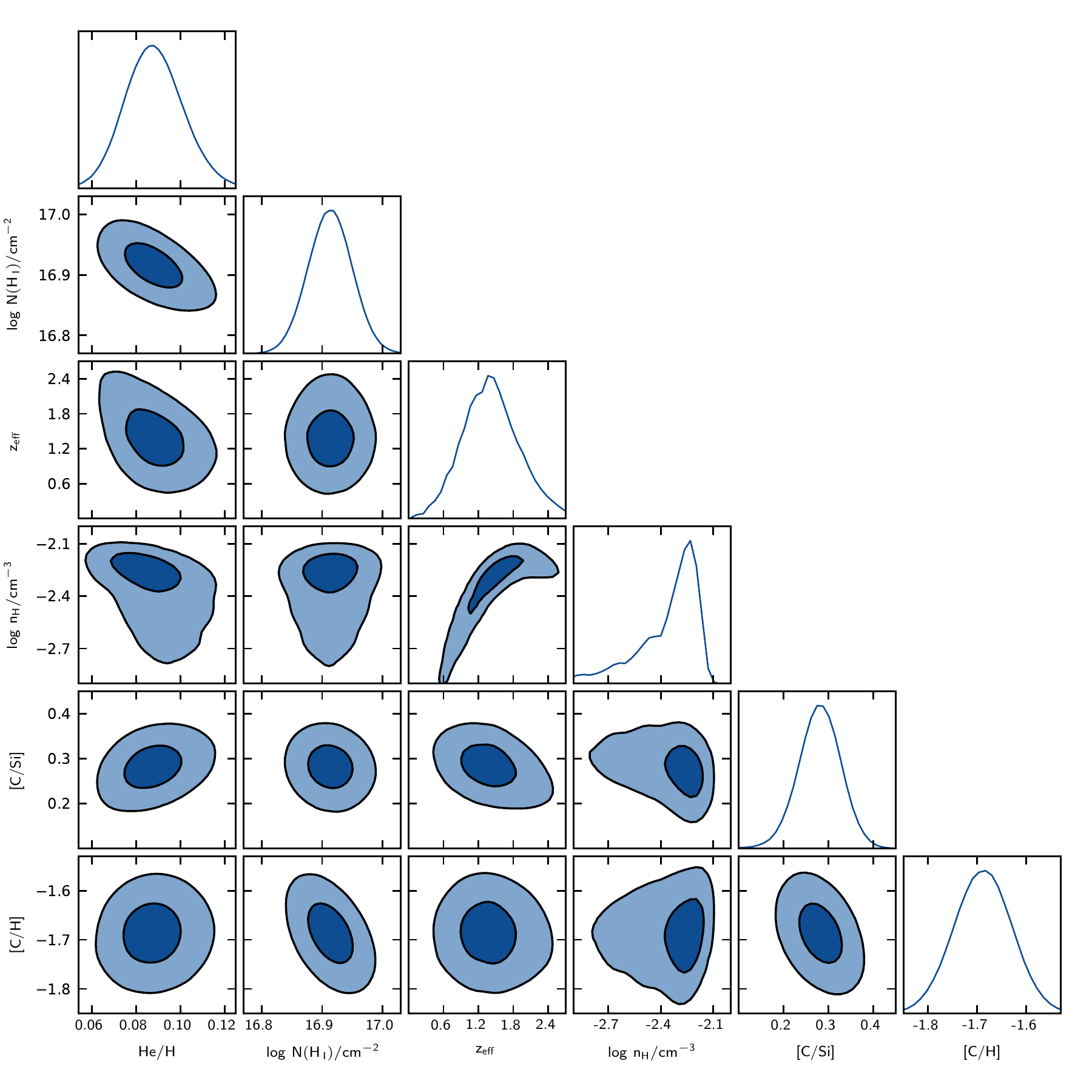}
\caption{\textbf{Confidence regions of the model parameters.} The 1D posterior distributions of each parameter are shown along the diagonal panels, while the off-diagonal panels show the two dimensional projections of each parameter combination, illustrating the covariance between model parameters. Dark and light shades represent the 68\%\ and 95\%\ confidence regions, respectively.}
\label{fig:mcmcfull}
\end{minipage}}
\end{figure}

\begin{figure}
  \centering
 {\includegraphics[angle=0,width=150mm]{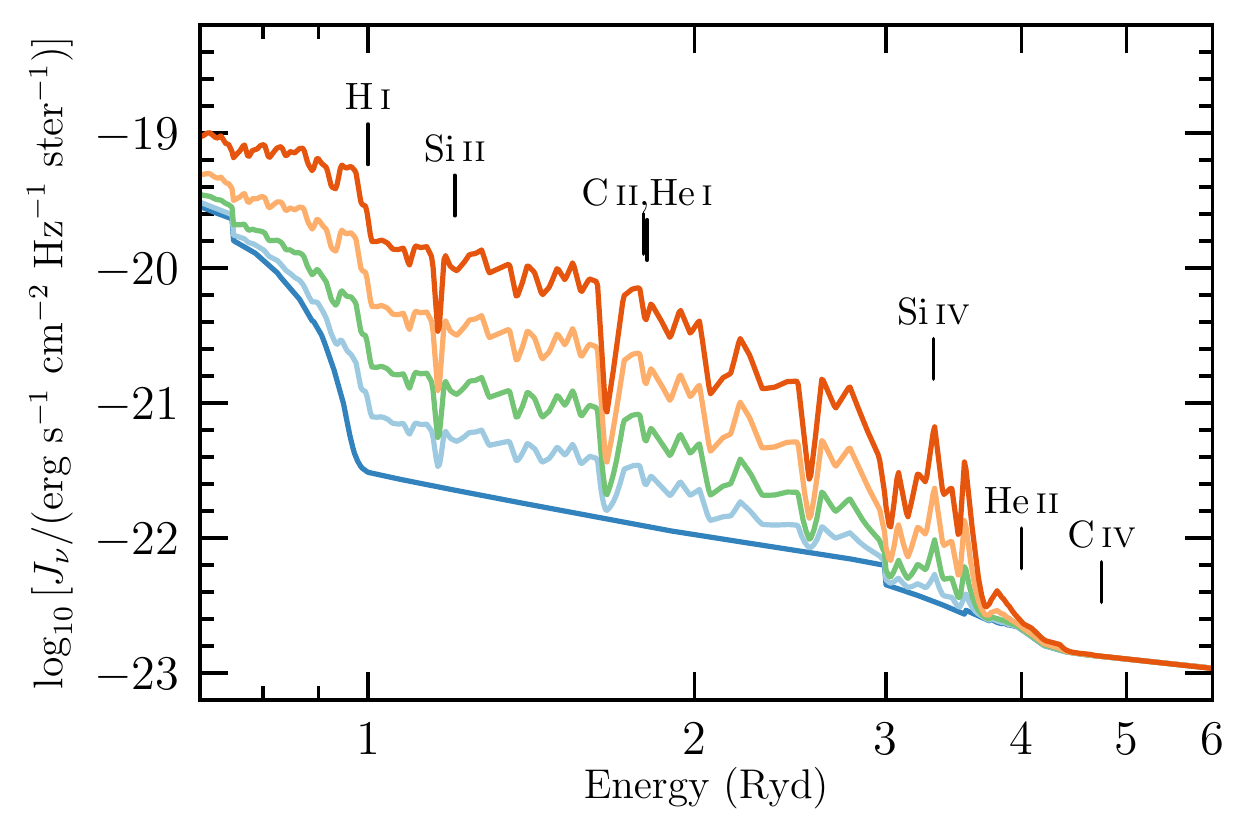}}\\
  \caption{
\textbf{Five sample radiation fields that include a contribution from the metagalactic UVB radiation field and a galactic radiation field.} The lowest blue curve shows the Haardt \&\ Madau UVB at \zabss, while the remaining curves (bottom to top) include an increasing contribution from a galactic radiation field (corresponding to $\log\,j_{\rm gal}=-\infty, -0.5, 0.0, +0.5, +1.0$). Labelled tick marks above the spectra indicate the ionisation potentials of the relevant ions that are used in this letter.
}
  \label{fig:uvbfgal}
\end{figure}

\begin{figure}
\noindent\makebox[\textwidth][c]{\begin{minipage}{183mm}
\includegraphics{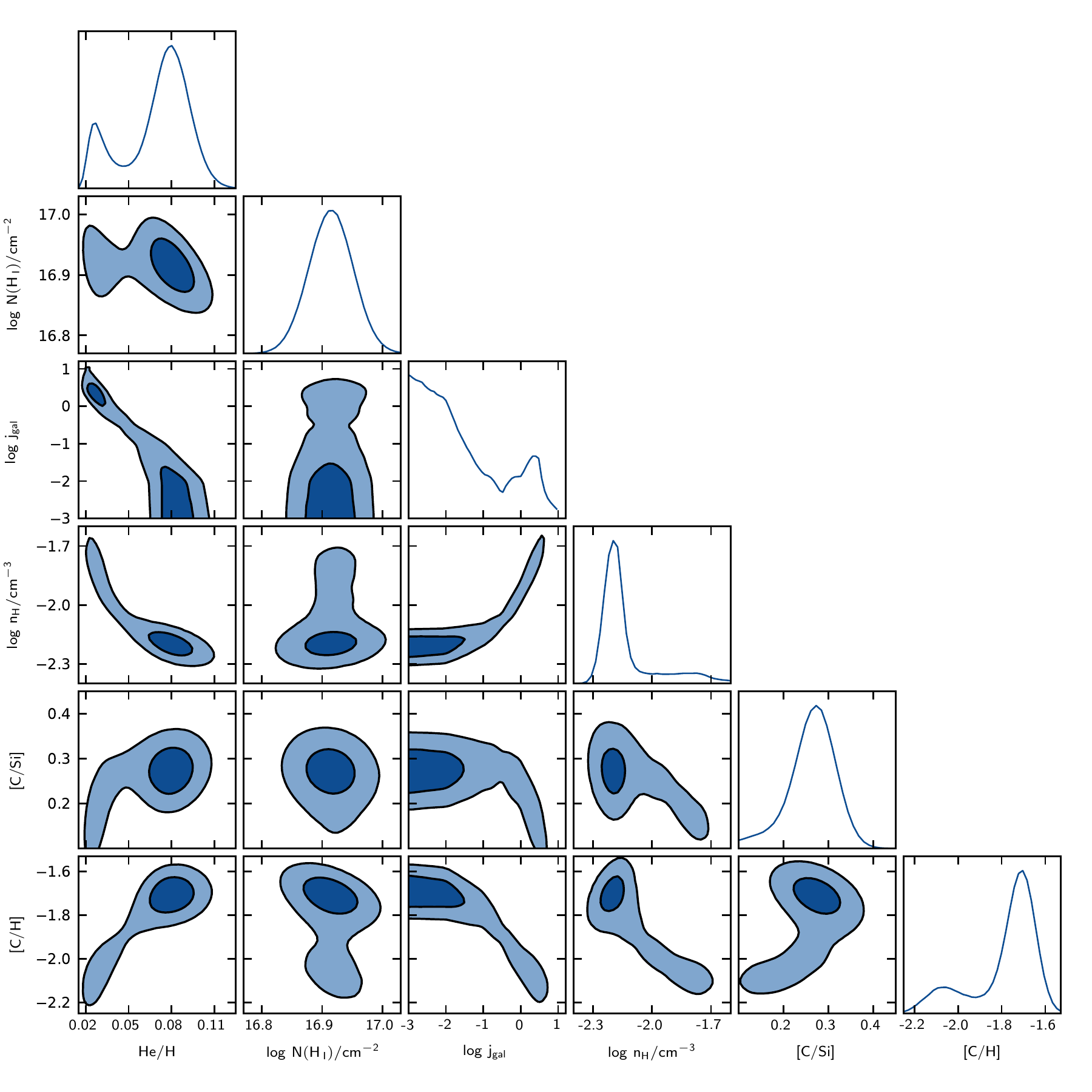}
\caption{\textbf{Confidence regions of the model parameters when using an incident radiation field that consists of a galaxy and the UVB.} The curves and contours have the same meaning as in Supplementary Figure~\ref{fig:mcmcfull}.}
\label{fig:mcmcjgal}
\end{minipage}}
\end{figure}


\end{document}